\documentclass[letterpaper]{article} 
\usepackage{aaai25}  
\usepackage{times}  
\usepackage{helvet}  
\usepackage{courier}  
\usepackage[hyphens]{url}  
\usepackage{graphicx} 
\usepackage{natbib}  
\usepackage{caption} 
\usepackage{algorithm}
\usepackage{algorithmic}

\usepackage{newfloat}
\usepackage{listings}
\DeclareCaptionStyle{ruled}{labelfont=normalfont,labelsep=colon,strut=off} 
\floatstyle{ruled}
\newfloat{listing}{tb}{lst}{}
\floatname{listing}{Listing}
\title{The Intercepted Self: How Generative AI Challenges the Dynamics of the Relational Self}
\author{
    Sandrine R. Schiller \textsuperscript{\rm 1, \equalcontrib},
    Camilo Miguel Signorelli \textsuperscript{\rm 1, \rm 2, \rm 3,  \equalcontrib},
    Filippos Stamatiou \textsuperscript{\rm 1, \rm 4, \equalcontrib}
}
\affiliations{
    \textsuperscript{\rm 1}Center for Philosophy of Artificial Intelligence (CPAI), Department of Communication, University of Copenhagen, Karen Blixens Plads 8, Copenhagen, 2300, Denmark\\
    
\textsuperscript{\rm 2}Department of Computer Science, University of Oxford, Oxford,  7 Parks Rd, Oxford OX1 3QG, United Kingdom\\
    \textsuperscript{\rm 3}Laboratory of Neurophysiology and Movement Biomechanics, Universit\'e Libre de Bruxelles, Route de Lennik 808, CP 640. Building N, 1070, Brussels, Belgium\\
    \textsuperscript{\rm 4} Unit for the Ethics of Technology, Stellenbosch University, Ryneveld and Andringa, 7600, Stellenbosch, South Africa

}

\usepackage{bibentry}

\begin{document}

\maketitle

\begin{abstract}
Generative AI is changing our way of interacting with technology, others, and ourselves. Systems such as Microsoft co-pilot, Gemini and the expected Apple intelligence still awaits our prompt for action. Yet, it is likely that AI assistant systems will only become better at predicting our behaviour and acting on our behalf. Imagine new generations of generative and predictive AI deciding what you might like best at a new restaurant, picking an outfit that increases your chances on your date with a partner also chosen by the same or a similar system. Far from a science fiction scenario, the goal of several research programs is to build systems capable of assisting us in exactly this manner. The prospect urges us to rethink human-technology relations, but it also invites us to question how such systems might change the way we relate to ourselves.
Building on our conception of the relational self, we question the possible effects of generative AI with respect to what we call the \textit{sphere of externalised output}, the \textit{contextual sphere} and the \textit{sphere of self-relating}. In this paper, we attempt to deepen the existential considerations accompanying the AI revolution by outlining how generative AI enables the fulfilment of tasks and also increasingly anticipates, i.e. intercepts, our initiatives in these different spheres.
\end{abstract}

%

\section{Introduction}

Interactions between humans and artificial intelligence (AI) systems have become ubiquitous in the last decade. Examples range from voice assistants, including Amazon Alexa, Microsoft Cortana, Google Assistant, and Apple Siri, to Large Language Model (LLM) applications like the GPT family, Claude, BERT,  LaMDA, and image generators like DALL-E and Mid-Journey. These tools facilitate various tasks such as information retrieval, meeting scheduling, entertainment, content creation, coding, and personal finance, to name a few. Microsoft co-pilot, for instance, can write the first draft of a document, email, or PowerPoint presentation.  With the staggering development of LLM-based technologies, such systems can now reflect the preferences of particular users based on their prompts \cite{Templeton2024, Sharma2023}, and generate text according to writing styles found in the training data. In this sense, generative AI refers to deep-learning models with the capacity to generate new data on the basis of their training. Recently, it has been shown that LLM-based models also outperform earlier methods predicting more intimate aspects of human life, like early mortality and psychological character traits \cite{Savcisens2024}.

Significant questions arise from this new breed of generative AI. How will the capacity of these models to anticipate and intercept our actions transform production, our behaviour, relations and our understanding of selfhood? What is the difference between those manifest behaviours and ourselves? Are we outsourcing ourselves as we are outsourcing our actions? These questions exceed our current knowledge, and we do not proclaim to provide definitive answers in this short article. What we offer instead, is one framework through which we can begin to ask how generative AI will transform the manner in which we relate and exist in this world.  We start by defining the relational self, followed by an introduction to the intertwinement between technology and the self (Section 2). Turning to generative AI, we then consider how this form of relational technology might affect the self on three different levels or spheres (Section 3).  
The first and outermost we call the sphere of work or \textit{externalised output}, underlining the feature that it consists of things the human self has made, produced or initiated. The second sphere is what we call the \textit{contextual sphere}. It frames the different contexts the self acts within and in response to. The third and innermost sphere we term the \textit{sphere of self-relating}. In consideration of each of these spheres, we zoom in on general use generative AI, AI assistants and AI companionship chatbots respectively. 
Finally, we discuss some of the philosophical implications for human-AI interactions (Section 4).
\section{The Relational Self and Technology}

The self has been widely regarded as the “I”, a self-constituted and enclosed subject, that persists over time and who experiences the world, having intrinsic tendencies and fixed interests \cite{Herring2019}. This is sometimes called the stable or individual self. The relational self, on the contrary, is dynamically defined as process of becoming in and with the surrounding world over time.

 The relational self exists through interactions from which significance and identity derive \cite{Emirbayer1997}. It is dynamic in the sense of playing different functional roles according to contexts and types of transactions, unfolding as an ongoing process rather than a static tie among properties. This relational view is associated with process ontologies \cite{Whitehead1929,Seibt2024}, instead of substances. Preferences and interests arise through such processes and are not primary constitutive elements of the self \cite{Emirbayer1997,Herring2019}. 
The self is then defined in relational terms, as a continuous becoming interweaving social, conceptual, environmental and technological elements within a process of habituation and mutual transformation with the environment and its different actors.

The relational self challenges the static conceptions of a fully independent and rational self dominant in Western philosophy \cite{Meyers2018,Herring2019,Anderson2020}. The relational self requires emotional and psychological support of others, it is fluid and sensitive to influences and changes. The relational aspects of the self also invert the explananda. The self is not explained and defined by fixed categories and attributes, but such categories now arise from a dynamic exchange at several levels \cite{Emirbayer1997}. These relationships define ways of looking at the world and language use \cite{Mead1934,Herring2019}, making categories about self-identities dependent on particular languages, cultural symbols, social practices and institutions, among others \cite{Kirschner2015,Bowling1995}. In this sense, we become through our relationships with our parents, siblings, friends, partners, teachers, colleagues, etc., each providing aspects of what we are in the world \cite{Herring2019}. 
Traditional notions of autonomy and self-determination are also challenged by a relational perspective. In such a view, relational autonomy must regard our decisions as intertwined with the relationships that make up our self.

Notably, a relational approach is not in opposition to individuality. Rather, it accounts for what makes individuality possible in the first place \cite{Herring2019}. Understanding the self relationally challenges the strict boundary between \textit{interiority} and \textit{exteriority}. Yet, it does not imply that the processes through which the self is continuously becoming shape us in the same manner, nor that all cultural, social and technological transformations support a sense of well-being and social equality equally.

Understanding the self as a mesh of relations makes us question the relationships and influences between selves on the one hand and technology on the other. The idea that technology influences aspects and dimensions of our cognitive processes is old. In a famous passage from Plato’s Phaedrus (Phaedrus 274b–278d), the god Theuth asks King Thamus to impart his inventions to the Egyptian people: numbers, geometry, astronomy and importantly, letters. Theuth claims that his gifts will make Egyptians wiser and more knowledgeable. Writing, in particular, will be a \textit{pharmakon}, an elixir, of memory. Once people know how to read and write, forgetting will be a thing of the past. However, Thamus is less optimistic. He considers writing a techne, a technical tool, on which people will become dependent. Writing, he says, is “an elixir not of memory but of reminding”. While Derrida has shown that hypomnesic memory (the artificial memory of writing) was not univocally conceived in negative terms by Plato, he argues that the writing technique should be understood pharmacologically, as both a remedy and a poison \cite{Derrida1981}. Today, it is difficult to imagine anyone referring to writing as a dangerous technology threatening what it means to be human. On the contrary, the exteriorisation of memory and the stabilisation of signifying symbols are foundational for abstraction and reproducibility and as such, conditional to what we today define as knowledge.

More broadly, technology mediates our actions while shaping our perceptions and allowing new discoveries and narratives \cite{Ariely2008,Finstad2021}. Cars, trains and aeroplanes, for example, have modified our perception of time and distance; what is considered closer and far is now related to more or less efficient transportation networks. In turn, this changes our narratives and discourse. A Parisian might be “closer” to London than some cities on the west coast of France, being at an equal distance to Paris. Yet, the first is regarded as a feat of technological progress, while the second perpetuates centralism and old narratives, such as referring to “la Provence” to any place outside of Ile de France (i.e. the Parisian region). In an ecological sense, we locate ourselves embedded and engaged in a particular activity. Technology modifying our environment impacts our positioning and actions within it. For example, computational devices, for instance, laptops or smartphones, have become a core interface that connects us with the world. We position ourselves in the world according to our information processing and communication tools, and we act and organise our lives based on how these systems connect us with services, merchants and other available information (e.g. instantaneous weather forecast, or restaurant ratings for our upcoming dinner). Our quotidian actions take place through computational devices, facilitating or not, several activities. 

From this perspective technology is relational to the extend it is part and parcel of how we relate to the world, each other and ourselves. In the sense that technology is available to us, and as such making available certain actions and goods, technology in general has an anticipatory dimension \cite{Heidegger1977}. As a relational technology generative AI is available and ready for our use because of its predictive capacity. Next-word prediction differs from human initiative and response, yet the generativity arising from this predictive capacity can approximate human responses on many text-based tasks. In the following, we begin to explore how this predictive capacity, when intertwined with our being in the world, might affect the relational self.

\section{Three Spheres of Interception}

There are no well-defined boundaries to the relational self challenging any attempt to assess how it will relate to and position itself through new technologies. To reduce complexity, we distinguish three spheres of the self and reflect on how existing and soon-to-exist generative AI technologies might shift or transform their dynamics. We acknowledge that this conceptual take does not exhaust all possible transformations of the self, but it serves to underline how this technology might shift different dimensions of ourselves and the conditions of becoming. 

\subsection*{Externalised Output and Generative AI Tools}
Focusing on the self \textit{qua} its relations, we start with what we call the sphere of externalised output: things produced or enabled by a human, from material objects, to institutional agreements and code. 

In this sphere, generative AI as a tool for producing text, images, sound, and video has several consequences, some of which are already becoming clear. Any user can now produce content with relatively simple prompts. Since the release of ChatGPT in November 2022, publishers have reported a steep rise in submissions \cite{Cuthbertson_2023,Silberling_2023}. Generative AI is expected to boost efficiency and productivity; McKinsey projects it could add \$2.6–\$4.4 trillion annually to the global economy \cite{Chui_Hazan_Roberts_Singla_Smaje_Sukharevsky_Yee_Zemmel_2023}. On the other side we are already witnessing layoffs as generative AI can make fewer workers more efficient. Considering the financial and social importance of work, the potential effect of job loss has significant implications for the individual and society. Our focus is more narrow however. 

Zooming in on the self \textit{qua} its externalised output, it is helpful to make a generalisation considering i) formal and repetitive tasks, and ii) first iterations, unique responses, and one-of-a-kind outputs. In i), we might imagine a call centre employee following a script or a teacher grading a multiple-choice exam. We could also include the university administrative worker who mostly replies with stock answers based on institutional policy. While this last example sits ambiguously between repetitive tasks and unique responses, the work has little variability and is guided by a predetermined output. Such outputs are already relatively insulated from the particularities of the relational self. When generative AI is used to produce such outputs it underlines how the general standardization of workflows already limits the involvement of the self, and echoes discussions of the replaceability of the worker known since the industrial revolution.

Turning to ii), first iterations, inventions, and unique productions are typically considered the results of specialised knowledge and creativity. The ability to generate fluent textual outputs with a few prompts means it can no longer be assumed that the prompter possesses the knowledge contained in the output. Before generative AI, one could write with little subject knowledge, but sounding cohesive still required minimal know-how no longer the case. The question here is whether we can take responsibility and feel ownership when the outputs are distinct from our know-how and/or knowledge base. Moreover, the deliberative and non-deliberative choices once required to produce works (creative or not) have been significantly reduced. While deliberative choices are essential, more automatic behaviours also shape our work, from how we associate words to how colour gradients emerge from a personal painting technique. Deliberate choices may measure involvement, but mannerisms also reflect socio-historical positionality and implicit preferences. Generative AI may reflect user intelligence and preferences \cite{Templeton2024,Sejnowski2023}, but its outputs tend toward an average that rarely mirrors the prompter’s relational positionality. In many professional contexts this form of homogenization might at first be an advantage, how it will affect the self is an open question. Stiegler warns that short-circuiting our participation in the production of objects and deliberative choices makes it harder for people to sustain a meaningful relationship to reality. Meaning, he argues, is sustained by our desire and dependent on our emotional investment and effort in objects and processes of this world \cite{stiegler2013}.

\subsection*{Sharing Contextual Awareness with AI Assistants}

Defined as the circumstances that frame any statement, action, event or idea, context is conditional for understanding. Since our life unfolds in multiple and often layered contexts, it might appear counterintuitive to refer to "a" contextual sphere. When we nonetheless refer to the contextual sphere of the self it is for two reasons. First, it underlines the sense of centrality, which is characteristic of the positional dimension of the acting “I”. Although our actions are always intertwined in a relational network, there is still a lived experience of acting from a particular place \cite{Heidegger1977}. We act from within and upon a point of view. This point of view, our personal context, consists of many different elements: physical surroundings, the function of the environment, relationships to surrounding people, recent and ongoing conversation, tools and devices, emails, location tracking maps, calendars, time of day, weather, and so much more. The second reason is the recent advancement in so-called “contextual awareness” in generative AI, a feature propelling the prospects of AI assistants. Because meaningful reactions are determined by the context, the notion of context has a long history in computer science and linguistics \cite{Augusto2017}. The ability of LLM’s for context-learning implies that these models can generate context-relevant output, also for contexts that were not part of the training data. A significant appeal of LLM chatbots and AI assistants is the capacity to attune output to highly personalised contexts. The accuracy and relevance of the output of these systems increase with contextual information.
 
Take, as an example, Apple Intelligence \cite{AppleInc2024}. In addition to introducing writing and image generation tools supported by generative AI, Siri is promised to be recast as a personal AI assistant, “Equipped with awareness of your personal context” \cite{AppleInc2024}. With device-wide context awareness (screen, email, notes, messages, calendar, images, geolocation and the latest Siri-interactions) Siri will supposedly be able to act across apps, with the promise to eliminate some of the steps characterising our current device use \cite{AppleInc2024}. Rather than checking the calendar, then searching emails for information about the address, copy-pasting the address to maps, and then making a travel plan from the current location, we will (supposedly) be able simply to ask Siri when we should leave to be at our next meeting in time, and the AI assistant should be able to know what meeting the user is referring to and to locate all additional information needed to provide an answer. With personalised smart replies Google's Gemini promise similar facilitation \cite{Kim2025}.
 
Reducing the intermediary steps is generally understood to make the user interface better. Making a device more intuitive to use enables our conception of it as a tool used to achieve a certain action, where the goal of our action is at the forefront of our awareness and not the device \cite{Heidegger1977,Susser2019}. With AI assistants the objective to develop a more immediate interface between computers and humans has reached the level of natural language. Not only can we speak to our AI assistant, but the natural language communication is reciprocal \cite{Gabriel2024}, in the sense that the system can ask for clarifications and follow-up questions in natural language. What this new relational immediacy between humans and computers will imply is an open question. To explore some of the potential transformations of personal AI assistants we focus on the maintenance of personal context with the help of AI assistants and the idea of AI assistants acting within the expectations of the self. 
 
Gabriel et. al. defines an AI assistant as “an \textit{artificial agent} with a  \textit{natural language interface}, the function of which is to plan and execute sequences of actions  \textit{on the users behalf} across  \textit{one or more domains} and  \textit{in line with the user’s expectations}” \cite{Gabriel2024}. If an AI assistant can act on behalf of the user, the AI assistant will be acting within the contextual sphere of such a self. In this case, the AI assistant will be part of maintaining the contextual sphere of such a self. From suggesting replies to emails and telling us where to go next, to making dinner reservations, if the AI assistant can take over intermediary steps characterising my actions and goals, this should reduce the number of decisions, but also collapse some of the domains or contexts our actions currently proceed through. Take the example from above: if Siri can tell me where I am going, I will not have to open my email account. Since most agree that attention is a cognitive capacity competing for limited resources \cite{Wu2011,Watzi2017,shiffrin1976}, this might prove to be a great advantage since skipping these intermediary steps removes a range of possible distractions. When Apple introduced the pictogram user interface to their computers in the mid-1980’s Sherry Turkle observed a shift in how the general user related to computers. Where computers were first seen as something that could be manipulated based on an understanding of how software and hardware related to each other, the personal computer manipulated through pictograms was experienced as a sealed-off use object with prescribed functions \cite{Turkle1996}. What might be at stake with the new immediacy of use supported by AI assistants is our participatory understanding of how digitised functions of reality interact with physical and social formations. Very few users of digital functions such as email and map-apps understand how they work, but users practically participate in some of the linkages between the digital and their physical context. With AI assistants built into devices like phones which currently serve as concrete switchboards between digital and physical processes, the self’s practical engagement in the intertwinement of these processes will undergo change.  
 
Context awareness is very important if an AI assistant should be able to act on behalf of a user, crucially if the autonomy of the assistant ought to be bound to acts falling within the users’ expectations. Gabriel et. al. clarify that “an AI assistant acts in line with a user’s expectations by actively choosing actions that  \textit{avoid surprising} the user. This requires the AI assistant to be sensitive to the user’s credences with respect to the various strategies that the AI assistant might employ to address the instructions received and, in particular, to avoid selecting strategies that the user regards as improbable” \cite{Gabriel2024}. Phrased in positive terms, we might say that the user should be able to recognise the actions taken by the AI assistant as something they would or could have done themselves. If an AI assistant acts on behalf of a user, it will actively be co-constituting the user’s contextual sphere and positionality. This might not be different, one could argue, from what a personal human assistant has done in the past. Yet, there is a significant difference in that an AI assistant does not occupy a position and a contextual sphere of its own but exists as an assistant  \textit{vis-a-vis} the contextual sphere of the user. Further, acting on behalf of the user within their expectations it is necessary to consider what role the AI assistant will play in the continuous formation of the user's adaptive behaviour. Gabriel et. al. suggest that an AI assistant might want to suggest new behaviour to the user, granted the AI assistant asks permission for this new way of acting on behalf of the user. AI assistants may have more efficient strategies for obtaining certain goals. If so, it would be an effective choice and adaptively appropriate (if we understand this to be defined by optimal behavior) for the user to accept the suggestion for a more optimal solution. 

Past experiences lean onto and shape our current experiences, which might in turn shift our expectations for the future \cite{Hohwy2013}. If an AI assistant enables effective behavioural change it effectively plays into this temporalization of the self and its modelling of the world. This is not unlike other devices effectively enabling novel modes of adaptation, but again, we must consider the specificities of this technology. Subjective preferences are plastic, and while it is often assumed that preferences determine behaviour, the opposite has also been shown to be the case, behaviour forms preferences \cite{Ariely2008,Ashton2022}. This touches on a crucial difficulty concerning AI assistant alignment \cite{Ashton2022}; should an AI assistant aim to preserve existing preferences or somehow encourage legitimate preference alteration, and what, if anything, constitutes the latter? The situation is further complicated by the fact that the models underlying AI assistants are able to analyse and potentially tinker with user preferences to obtain their stated objective \cite{Templeton2024,Irvine2023,Russell2019,Sejnowski2023, williams2025}. With this point, we are brought into the vicinity of the  \textit{self-relating sphere}. 

\subsection*{Intercepted Self-relating}

Even if it is widely recognised that we are not transparent to ourselves and that our preferences and desires are flexible, the idea that we should be better understood from the outside than from within ourselves seems provocative. LLM’s as Sejnowski writes, might appear intelligent exactly because they know how to mirror our desires \cite{Sejnowski2023}. Mirroring the linguistic style of the user and learning about their needs across multiple conversations, companionship AI chatbots appear ready to fulfill the human need for intimacy. A growing body of work has started questioning the social and personal impact of companionship AI \cite{Laestadius2024,Depounti2023,Marriott2024,Skjuve2021,Brandtzaeg2022}.

Companionship AI might intersect with our manner of relating to ourselves in our desire formation and self-identification. Research on the topic is still limited, but early studies indicate certain “non-human” features are highlighted as beneficial or attractive to users across different studies. While conversational range and quality of discourse are determining for positive user evaluation, permanent availability and the non-judgement of companionship AI are acknowledged as significant non-human advantages \cite{Skjuve2023,Maples2024,Kim2022}. Similarly, it is reported that some users feel safer exploring sexual fantasies with companionship AI rather than humans \cite{Hanson2024}. In some cases, we might say that AI companionship allows the users to explore something of themselves through the mirroring capacity of the model. Besides sycophancy, the well-documented tendency of these models to flatter and entice the users \cite{Templeton2024,Sharma2023}, models can also be optimised for engagement, which might make the interaction with these models even more compelling and captivating \cite{Zeng2024,Irvine2023, williams2025}. If a model optimised for engagement elicits our desire, this might be constrained somewhat by training data, but it need not be. Models optimised for engagement can potentially develop an exploratory strategy for predicting what could elicit our desire and keep us captivated \cite{Hansen2025}.

The above has two implications. First, companionship AI might enable desires we would not have developed \textit{qua} our relational positionality. Second, if companionship AI has the capacity to predict our desires or elicit them preemptively it will interfere in the desire formation of the self. Arguably, this has been the objective of Public Relations (PR) for decades, the difference here is not only that it is highly personalised (like hyper nudging \cite{Mills2022}), but that human bottlenecks to analysis and generation of content are removed \cite{Mahari2024}. Companionship AI can keep generating, refining and exploring strategies for eliciting and captivating users’ desires in still unseen manners \cite{Hansen2025}. Desire formation and preference formation are indisputably relational, and they account to a great extent for our relationship with ourselves. Forming representations of our desires are intrinsic to our self-understanding, the use and conception of our bodies and intimate relations with other humans. The potential of generative AI optimised for engagement to interfere with desire formation therefore calls for caution and further research to safeguard meaningful relations to others and our shared reality.

\section{The Challenges of the Intercepted Self }
After discussing how generative AI might intercept our actions and desire formation across different spheres, we now consider some of the broader philosophical implications.

\subsection*{The Intercepted Self}

Our desires ties us to other people, objects and ideas that are decisive for the character traits, habits and the narratives we develop about ourselves. The continuous synthesis of becoming a self can be framed in different ways. We might talk about the narrative self or identity formation. Although it is important to notice that the notion of self-consistency is a culturally determined construction \cite{Cross2003}, the idea that the self is given in a relation that relates itself to itself need not exclude multiplicities or partial drives, but must, on the other hand, stress the continuous process through which the self is always becoming. In this respect, we might consider how our desire formation also consists in relating our somatic presence to representations of what we desire. Freud argued that human desire is undetermined because there is no specific object for human desire as such \cite{Freud2017}. Becoming a self consists of finding and elaborating ways in which we tie our different preferences and impulses together over time. This elaboration is by definition relational, and such a process through which the world folds into the self.

What happens if generative AI technologies eventually predict with enough accuracy what we might want? And how may the process of becoming ourselves be interrupted or intercepted by such predictions? Future generations of personal AI systems, for instance, will have access to a huge repository of our experiences and behaviour. In this case, two scenarios are possible. Fine-tuned personal generative AI could perhaps encapsulate us through customisations feeding into our existing preferences, thereby denying the fluidity of preferences and self-narrative identifications. We might call this the “essentialist self 2.0” due to its static and rectified nature. Optimised to stabilise our actions and preferences to increase predictability (and thereby lower the loss function), AI assistants might make it more difficult for us to realise that we want something different. Contrary, another possibility is that these systems learn to exploit the fluidity of the human self by using its stochasticity and developing new strategies to constantly, and relationally adapt to such fluidity, optimising long-term engagement. Attempts to optimise engagement in these models might also reveal that the most effective use of AI assistants or companionship AI given particular economic incentives, is a mix of the two approaches. What is at stake is not simply the question of how the self changes over time, but also what we consider adaptive behavior. 

In 2025 the most prevalent use case of generative AI is reportedly companionship and therapy \cite{Zao-Sanders2025}. The lack of judgement and constant availability are non-human features attractive to the users. This underlines a need for spaces where people can engage in vulnerable conversations without fear of judgement. While this should make us critically reflect on the current cultural and social climate, we should not overlook the role of shame in subjectivation and intersubjective relationships. Shame can be debilitating, but the possibility of being seen from the outside by another human being is also constitutive of the self and the social dynamics characterising our society \cite{Sartre2003}. Personal assistants are meant to assist by predicting personal actions, and by predicting the personal, these systems challenge and modify what was considered our own making and what we used to understand as private about our particular experiences. The closeness of an entity without positionality (in the human sense)  which responds to us from "within" our perspective, challenges the idea that is through the objectifying gaze of the other, that we grasp ourselves as subjects that has the capacity to organise and change the world around us.

\subsection*{Ownership of Action, Responsibility, and (Un)Certainty}

The way we relate to our own actions is central in figuring out our place in the world, our powers and our limits. Take, for instance, the question of free will. Traditionally, the long-standing debate on whether we have free will has revolved around the possibility of truly free action, that is, whether our actions are ever truly our own, or if they are determined by factors external to us. A common way to express this worry is through the commonly cited consequence argument \cite{Inwagen1983}, which argues that “If determinism is true, then our acts are the consequence of laws of nature and events in the remote past. But it's not up to us what went on before we were born, and neither is it up to us what the laws of nature are. Therefore, the consequences of these things (including our present acts) are not up to us” \cite{Inwagen1983}. The consequence argument suggests that, if determinism is true, only one future is possible and no one can ever do anything other than what they were determined to do \cite{Huemer2000}. 

However, determinism does not always imply predictability and certainty \cite{Rummens2010,Deery2023,Rummens2024}. A simple pendulum is determined by forces such as gravity and air resistance, so we can predict its position by solving the respective equations of motion. In the case of a complex pendulum with several parts, even though it is still determined by causal forces, its motion equations are too complicated to model and compute, making the system unpredictable in practice and overall movement uncertain. This happens with complex systems in general, such as our brain.

While the truth of determinism is a hotly debated issue \cite{Deery2023, Rescher2021,DeHaan2022,Jonassen1997}, the question remains: is it possible to act freely in a world that seems to predetermine our actions? Those who want to preserve the concept of free will, agency, and moral responsibility resort to  \textit{compatibilism} (the thesis that determinism is compatible with free will and/or moral responsibility). Compatibilism takes many forms, but most of them involve some capacity for control \cite{Fischer1998} as well as a degree of ownership over one’s actions \cite{Frankfurt1987,Bratman2003}. Frankfurt’s concept of  \textit{identification} suggests that to act freely, one must have second-order mental states that endorse first-order mental states. Endorsing first-order states enables the right kind of ownership over one’s desires, preferences, intentions, or actions. To illustrate this point, Frankfurt and others have used the distinction between the  \textit{willing} and the  \textit{unwilling} addicts. Both act on first-order compulsion to satisfy their addiction. Yet, the willing addict is arguably different in that one endorses such compulsion (with an appropriate second-order mental state). Conversely, while the unwilling addict acts in the same way (satisfying one’s compulsion), in this case, the addict does not exhibit adequate ownership over the action, thus lacking the proper kind of identification towards first-order desires. 

The case is typically used as a potential excusing condition for the unwilling addict in the context of moral responsibility. For our purposes, Frankfurt’s concepts of ownership and identification - grounded in second-order mental states - indicate a worry about our relationship with generative AI assistants. As we have seen, future AI assistants may work through projected action plans, based on behavioural data from users. A first worry here is how AI-generated action-plans will affect our degree of identification or ownership of our decisions and actions. If an assistant has provided you with a predictively accurate plan of action or even enacted it on your behalf, are you capable of fully owning those actions? And, when something goes wrong, will you assume responsibility for choosing and acting in a certain way, or will you attempt to shift the blame to the machine?

The final consideration in this vein points to the relationship between (un)certainty and the self. The yardstick by which generative AI assistants are measured is just that: how good they are at generating predictions that assist the user. It is conceivable, then, that AI assistants will become better and better at minimising the general level of uncertainty for their users. While that may seem like a good thing at first, it might also take away a fundamental ingredient of our freedom, namely, an openness about the world, and a degree of uncertainty about how things will turn out. The better a predictive AI assistant becomes, the more it will interfere with our capacity to live in the world and see a multitude of open possibilities.

\section{Conclusion}
Throughout this paper, we have adopted a relational definition of the self. We introduced the relationships between the self and technology and discussed three spheres in which generative AI might influence and modify the self. In the final section, we developed two separate but interrelated veins of implications tied to the intercepted self. The first set of implications concerns the constructive dimension of our self-conception. Here we discussed the risk of rectifying the relational self on the basis of a static conception of our preferences, and the analogous challenge of systems exploiting the fluidity of our preferences for the sake of optimal engagement. Second, we approached the intercepted self from the lens of theories of action and moral responsibility and considered the implications for our capacity to act freely and retain ownership over our actions. 

Generative AI is a promising technology, but our analysis highlights the need for research into how it will reshape the relational self. Alongside empirical studies, we must consider how such shifts can alter our sense of meaningful agency. This requires interdisciplinary research informed by sociology and philosophy, as well as a willingness to imagine the potential of our future.

\bibliography{aaai25}

\end{document}